\documentstyle[twocolumn,amssymb,aps,epsfig]{revtex}
\tighten
\begin{document}
\draft

\title{\bf Carrier induced ferromagnetism in diluted magnetic semi-conductors.
}

\author{G.~Bouzerar and T.P.~Pareek}
\address{ 
Max Planck Institut f\"ur Mikrostrukturphysik\\
Weinberg 2,D--06120 Halle , Germany \\
 }
\address{~ 
\parbox{14cm}{\rm
\medskip
We present a theory for carrier induced ferromagnetism in diluted magnetic semi-conductor (DMS). Our approach treats on equal footing quantum fluctuations within the RPA approximation and disorder within CPA. This method allows for the calculation of $T_c$, magnetization and magnon spectrum as a function of hole, impurity concentration and temperature. It is shown that, sufficiently close to $T_c$, and within our decoupling scheme (Tyablicov type) the CPA for the itinerant electron gas reduces to  the Virtual Crystal Approximation. This allows, in the low impurity concentration and low density of carriers to provide analytical expression for $T_c$. For illustration, we consider the case of $Ga_{1-c}Mn_{c}As$ and compare our results with available experimental data.
\\ \vskip0.05cm \medskip PACS numbers: 71.30.Ds, 75.40.Gb, 75.50.Dd
}}
\maketitle

\narrowtext

The discovery of carrier induced ferromagnetism in DMS have attracted considerable attention from both
theoreticians and experimentalists. The interest for these material is
mainly stimulated by the possible technological applications
(e.g. semi-conductor spin devices). For example by doping GaAs\cite{Ohno1,Ohno2} with magnetic impurities $Mn^{2+}$, $T_{c}$ exceeding 100 K has been
reached. The doping of a III-V semiconductor compound with Mn
impurities introduces simultaneously local magnetic moments ($S=5/2$) and itinerant valence band carriers ($s=1/2$).
One of the important open issues is to find out whether it is possible
to reach critical Curie temperature of order 300 K.
Thus it is important to understand theoretically how $T_{c}$ varies with the impurity concentration, effective mass,
hole concentration and exchange integral.
Many theoretical approaches have been performed to analyze ferromagnetism in DMS, this includes mean field theory 
\cite{Dietl,Jungwirth,Bhatt}, spin wave theory \cite{Koenig1}, first principle calculations \cite{Sanvito,Akai,Shirai} and Monte-Carlo simulations \cite{Schlim}. In contrast to most of the theoretical work, we present a theory is able totreat disorder in a more realistic manner (beyond coarse graining). Our theory includes quantum fluctuations within RPA and disorder is treated within CPA. It should be stressed that in our approach the spin impurities are treated quantum mechanically.

We start with the following minimal Hamiltonian,

\begin{equation}
H=\sum_{ij, \sigma} t_{ij}c_{i\sigma}^{\dagger}c_{j\sigma}+
\sum_{i}J_{i} \vec{S}_{i}\vec{s}_{i} 
\end{equation}
The first term stands for the tight binding part of the itinerant free electron gas, $t_{ij}=t$ if i and j are nearest neighbor or 0 otherwise.
The second term is the exchange between localized impurities spin and
itinerant electron gas,
$J_{i}$ are random variables: $J_i=J$ if site i is occupied by a $Mn^{2+}$ ion or 0.
The operator $\vec{s}_{i}=c_{i\alpha}^{\dagger} (1/2 \vec{\sigma}_{\alpha \beta})c_{i\beta}$ is the spin operator at i of the itinerant electron gas and ${S}_{i}$ is the spin of the magnetic impurity.

Let us define the Green's function,

\begin{equation}
G^{+-}_{ij}(t)=-i\theta (t) \langle [S^{+}_{i}(t), S^{-}_{j}(0)] \rangle   =\ll S^{+}_{i};S^{-}_{j} \gg
\end{equation}

We write the equation of motion and use Tyablicov decoupling \cite{Tyablicov} (equivalent to RPA) which is suitable for ferromagnetic systems. It consists in closing the system by approximating the higher order Green's function
 $ \ll S_{i}^{z}s_{i}^+;S^{-}_{j} \gg \approx \langle S_{i}^{z}\rangle \ll s_{i}^+;S^{-}_{j} \gg $. In this approximation, we obtain in frequency space reads,

\begin{eqnarray}
(\omega+J_{i}\langle s^{z}\rangle  )G^{+-}_{ij}(\omega)= 2\langle S_{i}^{z}\rangle \delta_{ij} +
\nonumber \\
J_{i}\langle S_{i}^{z}\rangle \ll s_{i}^+;S^{-}_{j} \gg 
\label{propag1}
\end{eqnarray}
$\langle s^{z}\rangle$ is the magnetization of the itinerant
electron gas and $\langle S_{i}^{z}\rangle$ the magnetization of a
magnetic ion at site i. It is convenient to rewrite the new Green's function which appears in the right part of the equality in the following form,
$\ll s_{i}^+;S^{-}_{j} \gg =\frac{1}{L^2} \sum_{kq}
e^{iqR_{i}}\Gamma^{k+q,k}_{j}$, where $\Gamma^{k+q,k}_{j}=\ll
c_{k+q,\uparrow}^{\dagger}c_{k,\downarrow};S^{-}_{j} \gg$. We obtain,
\begin{eqnarray}
\Gamma^{k+q,k}_{j} = f(k,q,\omega) \sum_{l}\frac{1}{2}J_{l}e^{-iqR_{l}} G^{+-}_{lj}
\label{eqa}
\end{eqnarray}
where,
\begin{eqnarray}
 f(k,q,\omega)=\frac{(\langle n_{k+q,\uparrow}\rangle -\langle n_{k,\downarrow}\rangle )}{\omega-(\epsilon_{k} -\epsilon_{k+q})+
cJ\langle S_{A}^{z}\rangle }
\label{eqb}
\end{eqnarray}
$\langle n_{k,\sigma}\rangle$ is the occupation number of $(k,\sigma)$ state. 
c is the impurity concentration, $\langle S_{A}^{z}\rangle$ is the averaged magnetization of $Mn^{2+}$, 
and $\epsilon_{k}$ denotes the hole's dispersion.
Inserting both eq. (\ref{eqa}) and (\ref{eqb}) into eq. (\ref{propag1}) we immediately find,
\begin{eqnarray}
G^{+-}_{ij}=g_{i}\delta_{ij} + g_{i} \sum_{l} \phi_{il}G^{+-}_{lj}
\label{maineq}
\end{eqnarray}
where the T dependent locator $g_{i}$ is defined as,
\begin{eqnarray}
g_{i}(\omega)=\frac{2 \langle S^{z}_{i}\rangle }{\omega + J_{i} \langle s^{z}\rangle }
\end{eqnarray}
$\phi_{il}=\frac{1}{4} J_{i}J_{l} \chi^{0}_{il}(\omega)$ and $\chi^{0}_{il}(\omega)$ is the Fourier transform of the polarized susceptibility $\chi^{0}(q',\omega)$:
\begin{eqnarray}
\chi^{0}(q',\omega)=\frac{1}{L} \sum_{k}\frac{(\langle n_{k+q',\uparrow}\rangle -\langle n_{k,\downarrow}\rangle )}{\omega-(\epsilon_{k} -\epsilon_{k+q'})+
cJ\langle S_{A}^{z}\rangle }
\label{suscep}
\end{eqnarray}

Note that eq. (\ref{maineq}) {\it still contains the disorder} through $\phi_{il}$ and $g_{i}$.
It is also interesting to mention that the previous equation can be
interpreted as the propagator of a free particle moving on a
disordered medium, $g_{i}$ is the random on-site potential and
$\phi_{il}$ the long range-hopping terms. Note also that $\phi_{il}$
is energy dependent through $\chi^{0}_{il}(\omega)$. To solve the problem 
we have to calculate in a self-consistent manner $\langle s^{z}\rangle $ and $\langle n_{k,\sigma}\rangle $ which appear in eq. (\ref{maineq}). For that purpose we have to write the equation of
motion for the Green's function $ K_{ij,\sigma}=\ll
c_{i,\sigma};c^{\dagger}_{j,\sigma} \gg $. After decoupling we get,
\begin{eqnarray}
(\omega - \frac{1}{2}z_{\sigma}J_{i} \langle S^{z}_{i}\rangle ) K_{ij,\sigma} = 
\delta_{ij}+ \sum_{l} t_{il} K_{lj,\sigma}
\label{propa}
\end{eqnarray}

One can recognize the propagator of the Anderson model, with on-site random
potential depending on the spin $\sigma$:
$\epsilon_{i,\sigma}=\frac{1}{2} z_{\sigma} J_{i}\langle
S^{z}_{i}\rangle $. Since in our model the
potential is temperature dependent 
 through $\langle S^{z}_{i}\rangle $ then sufficiently close to
$T_{c}$ we will always be in the metallic regime $k_{f}l_{e}
\gg1$ \cite{Mesos}: $l_{e} \approx \frac{1}{(J\langle
S^{z}\rangle )^2}$. This is in contrast with the standard
Anderson model where the impurities are static.
Equations  (\ref{maineq}) and  (\ref{propa}) ($\sigma =\pm1$) provide a
closed system of equations which have to be solved self-consistently within CPA.

The simplest is to start with eq. (\ref{propa}). Indeed, it is straightforward to get the solution with the standard CPA since it contains only diagonal disorder. The averaged Green's function is

\begin{eqnarray}
K_{k,\sigma}=\frac{1}{\omega -\epsilon(k)-\Sigma_{\sigma}(\omega)}
\end{eqnarray}

where the self-energy is,
\begin{eqnarray}
\Sigma_{\sigma}(\omega)=V_{\sigma}-(\epsilon_{A,\sigma}-\Sigma_{\sigma}(\omega)) K_{\sigma}^{00}(\omega)(\epsilon_{B,\sigma}-\Sigma_{\sigma}(\omega))
\end{eqnarray}

where $\epsilon_{A,\sigma}=\frac{1}{2}z_{\sigma}J\langle S_{A}^{z}\rangle$, $\epsilon_{B,\sigma}=0$ and $V_{\sigma}$ is the average value $V_{\sigma}= \frac{1}{2}z_{\sigma}Jc\langle S_{A}^{z}\rangle $
and $K_{\sigma}^{00} =\frac{1}{L} \sum_{q}\frac{1}{\omega -\epsilon(q)-\Sigma_{\sigma}(\omega)}$.

The self-energy $\Sigma_{\sigma}(\omega)$ can re-expressed,

\begin{eqnarray}
\Sigma_{\sigma}(\omega)=V_{\sigma}[1+(1-c)\frac{1}{2}z_{\sigma}J\langle S_{A}^{z}\rangle K^{00}(\omega)]
\end{eqnarray}
We see that when $T \rightarrow T_{c}$,
$\Sigma_{\sigma}(\omega) \rightarrow \Sigma^{VCA}_{\sigma}(\omega)=V_{\sigma}$.
Thus, in the framework of our decoupling scheme, close enough to $T_{c}$ the CPA for eq. (10) reduces to the VCA. 

The final step of the calculation consists in solving 
eq.(\ref{maineq}). In order to provide analytical form for 
$T_{c}$ we use the similar approximation (VCA) for $G^{+-}_{ij}$ as done above for $K_{ij,\sigma}$. We expect this approximation to be reasonable in the limit of both low impurity concentration and low density of itinerant carriers. 
To get the averaged Green's function, we use the well-known Blackman-Esterling-Beck formalism \cite{BEB,Gonis}. By contrast with standard CPA, this approach 
is suitable for non diagonal disorder problems. It is based on a $2 \times 2$ matrix Green's function formalism for binary alloys using locator expansion.
Within VCA approximation, one gets for the averaged Green's function of an atom of type A,

\begin{eqnarray}
G^{+-}_{A}(k,\omega)=\frac{c}{g_{A}^{-1}-c\alpha(k,\omega)}
\end{eqnarray}

where $\alpha(k,\omega)=\frac{1}{4} J^{2} \chi^{0}(k,\omega)$.

Note also that since the Ga atoms have no magnetic moment, it implies $G^{+-}_{B}(k,\omega)=g_{B}=0.$
The $Mn^{2+}$ ion propagator can be rewritten,
$G^{+-}_{A}(k,\omega)=\frac{2}{E-E(q)}$, where $E=\frac{\omega}{\langle S_{z}\rangle }$. The dispersion $E(q)$ is solution of,

\begin{eqnarray} 
E(q)=-J\frac{\langle s_{z}\rangle }{\langle S_{z}\rangle } +\frac{1}{2}J^{2}\chi^{0}(q,E(q)\langle S_{z}\rangle )
\label{disp}
\end{eqnarray}

According to ref. \cite{Callen} the magnetization can be expressed in the following form,

\begin{eqnarray} 
\langle S_{A}^{z}\rangle=\frac{(s-\phi)(1-\phi)^{2S+1}+(S+1+\phi)\phi^{2S+1}}{(1+\phi)^{2S+1}-\phi^{2S+1}}
\end{eqnarray}
where $\phi=\frac{1}{L} \sum_{\bf q} \frac{1}{e^{\beta \omega(q)}-1} $.

When $T \rightarrow T_{c}$,
$ \phi= \frac{k_{B}T_{c}}{c\langle S_{A}^{z}\rangle} \frac{1}{L}\sum_{\bf q} \frac{1}{E(\bf q)}$. This implies for $T_{c}$ the standard RPA form,

\begin{eqnarray}
T_{c}=\frac{1}{3}c \frac{S(S+1)}{\frac{1}{N}\sum_{q}\frac{1}{E(q)}}
\label{eqtc}
\end{eqnarray}

This expression is similar to the one obtained in the clean limit for the Kondo Lattice Model \cite{Nolting,Yangetal}.
In the vicinity of $T_{c}$ the dispersion $E(q)$ is,

\begin{eqnarray}
E(q)=\frac{1}{8\pi^2} \frac{J^2}{t}\frac{1}{2} (k_{f} -\frac{1}{q}[k_{f}^2-\frac{q^{2}}{4}] 
ln(\frac{q+2k_{f}}{q-2k_{f}}))
\end{eqnarray}

Note that, below $T_{c}$, the eq. (\ref{disp}) should be solved numerically in order to get E(q) as a
function of the temperature. This is required to calculate $\langle S_{z}\rangle$ and $\langle s_{z}\rangle$ as function of T. According to eq. (\ref{eqtc}) $T_{c}$ is given by,
\begin{eqnarray}
T_{c}=\frac{S(S+1)}{24\pi^2} \frac{J^2
c}{t}(\frac{1}{N}\sum_{q}\frac{1}{C(q,k_{f})})^{-1}
\end{eqnarray}
where we define $C(q,k_{f})=\frac{1}{2} (k_{f} -\frac{1}{q}[k_{f}^2-\frac{q^{2}}{4}] 
ln(\frac{q+2k_{f}}{q-2k_{f}}))$. This implies that $T_{c}$ is proportional to $J^2$ and to the effective mass ($1/t$). 
The dependence on the hole concentration is only contained in $C(q,k_{f})$.
We define the hole concentration as $n_{h}=\gamma c$ where $\gamma \le 1$.
This is the simplest way to take into account the presence of As-antisites \cite{Sanvito2}.
In Fig. 1 we show the variation of $T_{c}$ as a function of $\gamma$.

\begin{figure}\centering
\vspace{-0cm}\hspace{-0cm}\epsfig{file=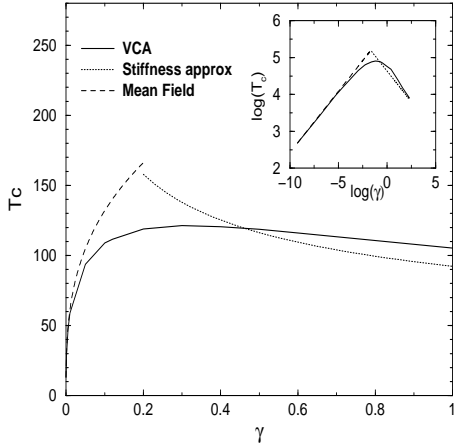,height=6cm,
width=6cm}\vspace{-0cm} \caption{$T_{c}$ as a function of $\gamma$ for $c=0.05$ and $\frac{J^2}{t}= 10.5 ~eV$.
 The continuous line represent the VCA calculation, the dashed line 
corresponds to $T_{c}$ within the mean-field approximation, and the dotted line is obtained by approximating
the dispersion by $E(q)=Dq^{2}$, D is the spin stiffness. The inset shows 
$log(T_{c})$ versus $log(\gamma)$.}
\label{fig1}
\end{figure}

We observe that in the low hole concentration regime, $T_{c}$ agrees very well with the mean field result (this is more clear in the inset log-log plot). In the mean field regime the magnon excitation spectrum is dispersionless: $E^{MF}(q)= \lim_{q \rightarrow \infty}
E(q)=\frac{1}{8\pi^2} \frac{J^{2}}{t} k_{f}$ where $k_{f}=(3\pi^2
\gamma c)^{1/3}$. In this limit,

\begin{eqnarray}
T_{c} =\frac{1}{24} (\frac{3}{\pi^4})^{1/3}S(S+1) \frac{J^{2}}{t} \gamma^{1/3} c^{4/3}
\end{eqnarray}

When increasing $\gamma$, $T_{c}$ {\it strongly} deviates from the mean
field results and shows a {\it broad} maximum. Such a maximum was also observed in ref. \cite{Koenig1}. By further increase of $\gamma$ the Curie temperature starts to decrease \cite{note1}. As we observe it from Fig. 1, for very large $\gamma$  $T_{c}$ agrees very well with the case where the magnon spectrum is approximated by
$E(q)=E^{stiff}(q)=Dq^{2}$ where the stiffness D is given by $D=\frac{1}{48 \pi^2} \frac{J^{2}}{t}
\frac{1}{kf}$, this regime is denoted ``stiffness'' regime. In this regime we find,

\begin{eqnarray}
T_{c} =\frac{1}{144} \frac{1}{(18\pi^4)^{1/3}} S(S+1) \frac{J^{2}}{t} \gamma^{-1/3} c^{2/3}
\end{eqnarray}

The existence of a maximum can be understood in the
following way: Like in the RKKY situation \cite{rkky}, the exchange oscillate with typically 
length scale $l_{osc} \propto 1/k_{f}$. Thus it is expected that when the
length scale gets sufficiently large (larger than the average distance
between impurities) some Mn-Mn bonds are coupled
antiferromagnetically. The induced frustration has for immediate consequence a
decrease of $T_{c}$. In Fig. 2 we illustrate the previous discussion by showing the dispersion as a function of $k/k_{c}$ where $k_{c}$ is chosen in order to conserve the volume of the Brillouin zone $(v=(2 \pi)^{3})$.
The results are shown for the 3 different regions: ``mean field'',
``intermediate'' and ''stiffness'' regime. We observe that in all 
cases the dispersion goes to 0 (when $q \rightarrow 0$), as expected when the Goldstone theorem is fulfilled.

\begin{figure}\centering
\vspace{-0cm}\hspace{-0cm}\epsfig{file=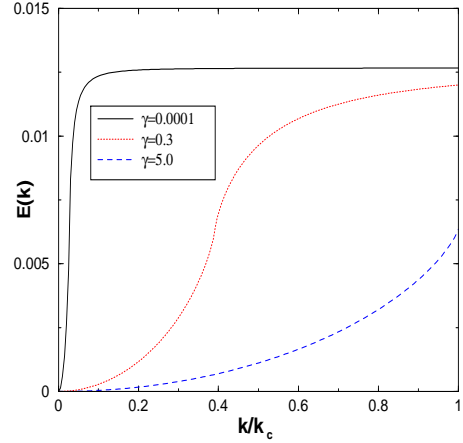,height=6cm,
width=6cm}\vspace{-0cm} \caption{
Magnon dispersion for $c=0.05$ in the 3 different regime, ``mean field'' (continuous line), ``intermediate'' (dotted line) and ``stiffness'' regime (dashed line). $E(q)$ is rescaled by a factor $\lambda=\frac{J^2}{t}(3\pi^2\gamma)^{1/3}$
}
\label{fig2}
\end{figure}

In Fig. 3, we show the region for which $T_{c}$ reaches 
its maximum as a function of c ($T_{c}^{max}(c)$) and the region where MF
formula provides a good approximation for $T_{c}$, it corresponds to 
$| \frac{T_{c}-T_{c}^{MF}}{T_{c}} | \leq 0.1$.
First we see that the region of validity of the MF result (dashed
area) corresponds to a very narrow region typically $\gamma \leq
0.05$. A good approximated value of the $\gamma$ for which $T_{c}$ is
maximum can be obtained by taking the intersection point between the
MF and ``stiffness'' values. This leads to $\gamma_{max} c=n_{max}=0.016$.

\begin{figure}\centering
\vspace{-0cm}\hspace{-0cm}\epsfig{file=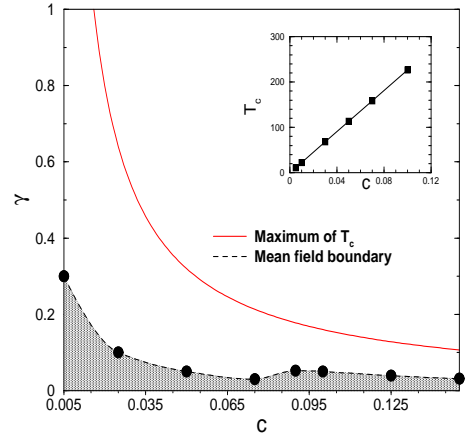,height=6cm,
width=6cm}\vspace{-0cm} \caption{
The dashed area represents region where mean-field result for $T_{c}$ is valid, the symbols are calculated points and the dashed line a fit.
The continuous curve represents values of ($\gamma$,c) for which $T_{c}$ is maximum ($T_{c}^{max}$) .
In the inset we have plotted  $T_{c}^{max}$ as a function of c assuming $\frac{J^2}{t}= 10.5 ~eV$  
}
\label{fig3}
\end{figure}

So far, for our discussion we did not have to specify the values of the parameters $t$ and
$J$. In order to check the validity of our theory we compare our results 
with available experimental data. GaAs is known to have a fcc
structure with a lattice constant $a_{0} \approx 5.6 A$. For
simplification in our calculation we have assumed a simple cubic
structure thus the lattice constant which has to be taken in our
calculation is $a_{1}=\frac{a_{0}}{4^{1/3}}$ in order to conserve the
volume for the unit cell. By also assuming an effective mass for the holes $m=0.5~ m_{e}$ one gets $t=0.63 ~eV$. The remaining free parameter $J$ will be chosen in order to fit the experimental data of ref.\cite{Ohno2}. For that purpose we calculate $\gamma$ for each sample according to the measured experimental values of the hole concentration given in Fig. 2 of ref.\cite{Ohno2}. 
The results are depicted in Fig. 4. As it can be seen we find a very
good agreement with the experimental data if $\frac{J^{2}}{t}=10.5
~eV$, this implies $|J|= 2.58 ~eV$ \cite{note2}.
Note that the deviations observed at low c are due to the uncertainty on the
hole concentration value (see the huge error bars in Fig. 2 of ref.\cite{Ohno2}).

\begin{figure}\centering
\vspace{-0cm}\hspace{-0cm}\epsfig{file=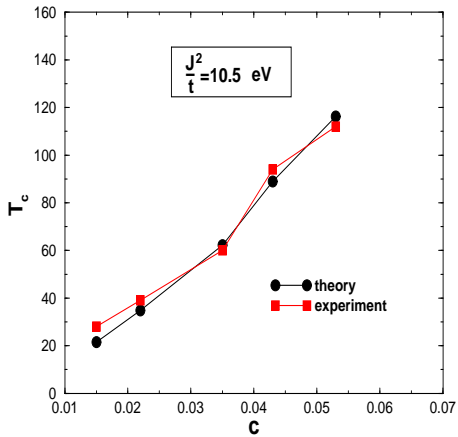,height=6cm,width=6cm}
\vspace{-0cm}\caption{
$T_{c}$ in Kelvin as function of c.
The full square corresponds to experimental values taken from ref.\protect\cite{Ohno2}.
The full circles represent the calculated values, the $\gamma$'s where also taken from the same reference.}
\label{fig4}
\end{figure}

From the experimental measurements, there is no clear consensus concerning the correct value of this parameter. Indeed, recent core level photoemission has provided $J=-1.2 \pm 0.2 ~eV$ \cite{Okabayashi}.
Whilst, from Magneto-transport measurements a value of $|J|=2.4 \pm 0.9 ~eV$ was suggested.
\cite{Ohno2,Omiya}. And within first principle calculations Sanvito et al.\cite{Sanvito} have found $J \approx -4.65 \pm 0.25 ~eV$.
In order to proceed to a better estimation of the parameters J one should compare theoretical calculations with other data, for instance transport measurements data \cite{Bouzerar2}. However, it is interesting to note that the band splitting at $T=0 K$ ($\Delta = J c S$) obtained within our calculations agrees with the experimental value reasonably well \cite{Szczytko}.
In the inset of Fig. 3, assuming $\frac{J^{2}}{t}=10.5~eV$ we show
$T_{c}$ as function of c taking $\gamma$ on the line of ``maximum of
$T_{c}$''. For instance if $c \approx 0.1$ and $\gamma \approx 0.2$ a $T_{c}$ of order 230 K can be reached.

\begin{figure}\centering
\vspace{-0cm}\hspace{-0cm}\epsfig{file=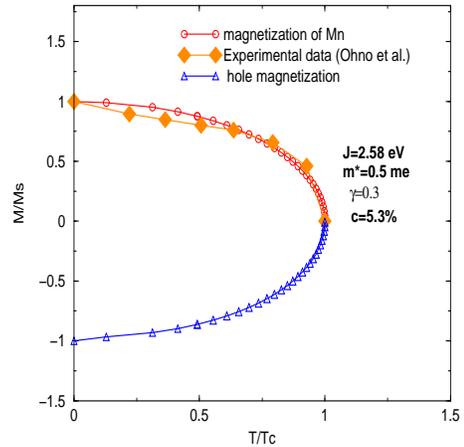,height=6cm,width=6cm}
\vspace{-0cm}\caption{Normalized magnetization as function of $T/T_{c}$. The experimental data are taken from \protect\cite{Ohno2}.
}
\label{fig5}
\end{figure}

Let us proceed further on and compare the calculated magnetization with the measured one. In the experimental data the concentration of $Mn^{2+}$ is $c = 0.053 \%$ and the parameter $\gamma$ is estimated to be 0.3 (see ref. \cite{Omiya})
In Fig. 5, we show the magnetization as function of the T where $m^*=0.5~ m_{e}$ and $J=2.58 ~eV$. .
We observe that for sufficiently high temperature $T \le 0.5 ~T_{c}$ there is a very good agreement with the measured $Mn^{2+}$ magnetization. When decreasing T some deviation appears, this suggests that the VCA treatment is not good enough in this region, which was expected.

To conclude, we have presented a general theory for carrier induced ferromagnetism in DMS. Our approach allows to treat the disorder beyond simple coarse graining within  full CPA treatment.
It goes beyond mean-field and includes quantum fluctuations in the RPA approximation. We have shown that, within our decoupling scheme and sufficiently close to $T_{c}$ the CPA for the itinerant gas reduces to VCA, which allowed us to provide analytical results for $T_c$ in the low impurity concentration and hole density regime. We have also discussed its dependence on the hole concentration. We have also shown that the mean field approximation is only valid for very low carrier concentration. Additionally, for illustration of our theory a comparison with available experimental data on $Ga_{1-c}Mn_{c}As$ was done. We find a very good agreement with the experimental results assuming a single band for itinerant carriers and a large exchange constant $J=-2.58  ~eV$.
Finally, this work provide a good starting point for higher decoupling scheme.

{\bf Note added:} After this work was completed we became aware of Yang et al. comment\cite{Yangetal}. By analogy with the Kondo Lattice Model \cite{Nolting} (no disorder) they proposed a similar expression for $T_{c}$ to the one derived in eq.(\ref{eqtc}).


\begin{references}

\bibitem{Ohno1} H. Ohno, H. Munekata, T. Penney, S. von Molnar, and L.L. Chang, Phys. Rev. Lett.
{\bf 68}, 2664 (1992).
\bibitem{Ohno2} F. Matsukura, H. Ohno, A. Shen, and Y. Sugawara, Phys. Rev. B{\bf 57}, R2037 (1998).
\bibitem{Dietl} T. Dietl, A. Haury, and Y.M. d'Aubigne, Phys. Rev. B{\bf 55}, R3347 (1997);T. Dietl, H. Ohno, and F. Matsukura cond-mat/007190.
\bibitem{Jungwirth} T. Jungwirth, W.A. Atkinson, B.H. Lee, A.H. MacDonald, Phys. Rev. B{\bf 59}, 9818 (1999); B.H. Lee, T. Jungwirth, A.H. MacDonald, Phys. Rev. B{\bf 61}, 15606 (2000).
\bibitem{Bhatt} R.N. Bhatt and M. Berciu cond-mat 0011319.
\bibitem{Koenig1} J. K\"onig, H.H. Lin and A.H. MacDonald, Phys. Rev. Lett. {\bf 84}, 5628 (2000);
J. K\"onig, T. Jungwirth, and A.H. MacDonald cond-mat/0103116.
\bibitem{Sanvito} S. Sanvito, P. Ordej\'on and N.A. Hill, Phys. Rev. B {\bf 63}, 165206 (2001);
\bibitem{Akai}H. Akai, Phys. Rev. Lett. {\bf 81}, 3002 (1998).
\bibitem{Shirai}M. Shirai, T. Ogawa,, I. Kitagawa,and N. Suzuki, J. Magn. Mater. {\bf 177-181}, 1383 (1998).
\bibitem{Schlim} J. Schliemann, J. K\"onig, and A.H. MacDonald cond-mat/0012233.
\bibitem{Tyablicov} S.V. Tyablicov, {\it Methods in quantum theory of magnetism} (Plenum Press, New York, 1967).
\bibitem{Mesos} P. W. Anderson Phys . Rev {\bf 102} 1008 (1958), B. L. Altshuler and A. G. Aronov in ``Electron-Electron interaction in disordered systems'', Edited by A. F. Efros and M. Pollack, North Holland (1985).
\bibitem{BEB} J. A. Blackman, D. M. Esterling, and N. F. Beck, Phys. Rev. B {\bf 4}, 2412 (1971).
\bibitem{Gonis} A. Gonis and J. W. Garland, Phys. Rev. B {\bf 16}, 1495
(1977).
\bibitem{Callen} H. B. Callen Phys. Rev. {\bf 63} 890 (1963).
\bibitem{Nolting} W. Nolting, S. Rex and S. Mathi Jaya, J. Phys. condens.
matter {\bf 9}, 1301 (1997).
\bibitem{Yangetal} M. F. Yang, S. J. Sun and M. C. Chang Phys. Rev. Lett. {\bf 86}, 5636 (2001).
\bibitem{Sanvito2} S. Sanvito and N.A. Hill Applied Physics Lett.,
 {\bf 78}, 3493 (2001).
\bibitem{note1} The expression of $T_{c}$ for $\gamma \gg 1$ is 
not expected to be valid since the VCA should break down (spin glass phase). This is not the regime in which we are interested in.
\bibitem{rkky}C. Kittel, in {\it Solid State Physics}, edited by
F. Seitz, D. Turnbull, and H. Ehrenreich (Academic, new York, 1968),
vol. 22, p.1 .
\bibitem{Omiya}T. Omiya, F. Matsukura, T. Dietl, Y. Ohno, T. Sakon, M. Motokawa and H. Ohno,
Physica E {\bf 7},976 (2000).
\bibitem{note2} We believe that beyond a single band treatment of the itinerant carrier one should find a smaller value of J. (see ref.\cite{Koenig1} cond-mat 0103116)
\bibitem{Okabayashi}J. Okabayashi, A. Kimura, O. Rader, T. Mizokawa, A. Fujimori, T. Hayashi,
and M. Tanaka, Phys. Rev. B {\bf 58}, R4211 (1998).
\bibitem{Bouzerar2} G. Bouzerar in preparation.
\bibitem{Szczytko} J. Szczytko et al. Phys. Rev. B  {\bf 59}, 12935 (1999).
\end{references}
\end{document}